\documentclass[fleqn,usenatbib]{mnras}

\usepackage{newtxtext,newtxmath}
\usepackage[T1]{fontenc}
\DeclareRobustCommand{\VAN}[3]{#2}
\let\VANthebibliography\thebibliography
\def\thebibliography{\DeclareRobustCommand{\VAN}[3]{##3}\VANthebibliography}

\usepackage{amsmath}	% Advanced maths commands
\usepackage{hyperref}
\usepackage{xurl}
\usepackage[pdftex]{graphicx}
\newcommand{\as}{$^{\prime\prime}$}
\newcommand {\pry}[1]{#1}
\newcommand{\kms}{km~s$^{-1}$}

\title[CLV in CH and QS]{Center-to-limb variations in coronal hole and quiet Sun regions obtained with IRIS spectroscopic observations}

\author[Kayshap $\&$ Young]{
Pradeep Kayshap,$^{1}$\thanks{E-mail: virat.com@gmail.com}
Peter R. Young$^{2,3}$
\\
$^{1}$ School of Advanced Sciences and Languages, VIT Bhopal University, Kothrikalan, Sehore Madhya Pradesh - 466114\\
$^{2}$ NASA Goddard Space Flight Center, Greenbelt, MD, USA\\
$^{3}$ Department of Mathematics, Physics and Electrical Engineering, Northumbria University, Newcastle upon Tyne, UK
}

\date{Accepted XXX. Received YYY; in original form ZZZ}

\pubyear{2023}

\begin{document}
\label{firstpage}
\pagerange{\pageref{firstpage}--\pageref{lastpage}}
\maketitle

\begin{abstract}
The center-to-limb variations (CLV) of Gaussian fit parameters of the  transition region Si~{\sc iv} 1402.77~{\AA} spectral line in quiet Sun (QS) and coronal hole (CH) regions are presented. The results are derived  from a full-disk mosaic scan obtained by the Interface Region Imaging Spectrograph on 24 September 2017. The CLV for a CH transition region line has not previously been reported, and the parameters are found to show variations consistent with the QS. The intensity increases towards the limb, consistent with an increasing plasma column depth due to line-of-sight effects. The Doppler velocity is normalized to be zero at the limb for both QS and CH and increases to $+4.8$~\kms\ (redshift) at disk center for CH and $+5.2$~\kms\ for QS. Non-thermal broadening in the CH decreases from a maximum of 24~\kms\ at the limb to 10~\kms\ at disk center. For QS the broadening decreases from 25~\kms\ at the limb to 14~\kms\ at disk center. Both Doppler velocities and non-thermal velocities vary linearly with $\cos\,\theta$, where $\theta$ is the heliocentric angle. The QS results for both parameters are consistent with earlier work.

%The work aims to understand the center-to-limb variations (CLV) in Quiet-Sun (QS) and coronal holes (CHs) using transition-region (TR) Si~{\sc iv} 1402.77~{\AA} spectral line observed by IRIS. The CLVs are present in the spectroscopic parameters (i.e., intensity, Doppler velocity, and non-thermal velocity) of QS as well as CH.  The Doppler velocity and non-thermal velocity vary linearly with $\mu$ in both regions (i.e., QS and CH). While the intensity does not vary linearly. This work reports CLVs in CH spectroscopic parameters for the first time ever. 
\end{abstract}
    
\begin{keywords}
Sun: transition region -- Sun: atmosphere -- Sun: UV radiation
\end{keywords}

%%%%%%%%%%%%%%%%%%%%%%%%%%%%%%%%%%%%%%%%%%%%%%%%%%

%%%%%%%%%%%%%%%%% BODY OF PAPER %%%%%%%%%%%%%%%%%%
\section{Introduction}
Ultraviolet emission lines formed in the solar transition region (TR) at temperatures of $10^5$~K have long been known to exhibit Doppler redshifts and broadening in excess of the thermal broadening. \citet{1976ApJ...205L.177D} were the first to report redshifts in TR lines from a study of \emph{Skylab} S082-B spectra, finding Doppler shifts of up to 15~\kms. Enhanced non-thermal broadening at TR temperatures around $10^5$~K was first reported by \citet{1973A&A....22..161B}. Both results likely give clues on the how mass and energy are balanced in the solar atmosphere. For example, non-thermal broadening may be caused by the passage of magnetohydrodynamic waves through the TR, while ubiquitous downflows in the TR represent mass and energy loss terms that need to be balanced to maintain a hot corona \citep[see][for more details]{1992str..book.....M}.
%For example, enhanced line broadening may indicate Alfv\'en waves for magnetoacoustic waves transporting energy to the corona, while systematic Doppler shifts may reveal 
%, and have been the subject of a number of modeling studies over the past five decades \citep{2010ApJ...718.1070H}

One method for characterizing the TR non-thermal broadening and Doppler shifts  is to measure the emission line properties at different positions on the Sun, from the center of the disk to the limb. Variations in the parameters then give clues as to the origin of the two effects. For example, if plasma flows are along radially-aligned structures then net Doppler shifts would be expected to be at a maximum at disk center and fall to zero at the limb. Similarly, if non-thermal broadening is due to random, lateral motions of radially-aligned structures then it would be expected to be largest at the limb and fall to zero at disk center (the motions giving a distribution of Doppler shifts when the line-of-sight is aligned to the motions).

In the present work we present the center-to-limb variation (CLV) of the Doppler velocity and non-thermal broadening of the \ion{Si}{iv} 1402.77~\AA\ TR emission line in both quiet Sun and coronal hole conditions across the full range of heliocentric angles. Although such a study has been performed for the quiet Sun previously, the present work is the first for coronal holes. \ion{Si}{iv} is formed at around 80~kK in the solar atmosphere, and the earlier work of \citet{PeterJudge1999}  suggests this line should show a redshift of 5--8~\kms\ at disk center in the quiet Sun.

CLV of emission line properties has been studied in various ways with spectrometers in the past. The HRTS rocket experiment had a long slit that could be positioned to extend from disk center to the limb. \citet{1984ApJ...281..870D} studied the CLV of several lines, including the \ion{C}{iv} 1548~\AA\, 1550~\AA\ doublet, and \citet{Brekke1993} looked at the CLV of two \ion{O}{v} lines. Another rocket experiment performed an east-west scan across the equator, with results for several lines presented in \citet{1990ApJ...358..693R} and \citet{1991ApJ...372..710H}. Spacecraft enable long-duration observations and full-disk raster scans were performed with the Solar Ultraviolet Measurements of Emitted Radiation \citep[SUMER:][]{1995SoPh..162..189W}. These have been used to study the CLV of several lines \citep{Peter1999}. These studies all yield continuous data from disk center to the limb, either with a single observation or with a patchwork of multiple spacecraft pointings. As spacecraft spectrometers usually have a restricted field-of-view, authors sometimes combine observations obtained at different times and pointings to investigate CLV \citep{1976ApJ...205L.177D,Rao2022}. In the present work we use IRIS data obtained with a full disk scan, analogous to the study of \citet{Peter1999}.

For studying the Doppler shift CLV, an important consideration is the reference wavelength used to derive the line-of-sight velocity. Within a single spectrum a common procedure is to assume a chromospheric emission line  within the spectrum has a zero Doppler velocity, which then fixes the absolute calibration of that spectrum. The measured wavelength of the TR line is then compared with the known rest wavelength to yield the Doppler velocity. Tables of rest wavelengths are given in \citet{Brekke1997} and \citet{1987JPCRD..16S....K}. The TR Doppler velocities would be expected to go to zero at the limb, and this was validated by \citet{1990ApJ...358..693R} and \citet{PeterJudge1999}. The earlier work of \citet{1984ApJ...281..870D} suggested this may not be the case, but \citet{Brekke1997} argued that a single slit position \citep[such as used for the][observation]{1984ApJ...281..870D} may lead to non-zero Doppler shifts at the limb, but a spatial average (multiple slit positions) will lead to an average zero Doppler velocity. 

\citet{PeterJudge1999} provided  quiet Sun Doppler velocity CLV data for 10 ion species, fitting the results with a $\cos\,\theta$ function ($\theta$ the heliocentric angle). Only the spatial region 750\as\ to 1000\as\ was fit and so the disk center velocity was inferred from the $\cos\,\theta$ form of the fit. The results were obtained from a special observation where the SOHO spacecraft was progressively rolled through 30$^\circ$ steps with the SUMER slit positioned 70\as\ inside the limb. Some coronal hole data were available from this dataset, and \citet{PeterJudge1999} provided CLV curves for eight ion species. The spatial region that could be fit was smaller than for the quiet Sun, and varied with ion species. The authors noted that the ion \ion{C}{iv}, formed at a similar temperature to the \ion{Si}{iv} considered in the present work, showed a CLV behavior significantly different to the quiet Sun. Inspection of the authors' Figure~5 shows Doppler velocities that are slightly positive compared to $-1$ to $-2$~\kms\ in the quiet Sun. 

TR emission line widths have generally been found to be approximately constant from disk center to the limb, with a small increase at the limb \citep{1976ApJ...209..270F,1998ApJ...505..957C}. The limb increase found by \citet{2000A&A...356..335D} for \ion{C}{iv} was interpreted as due to opacity effects at the limb. \citet{1998A&A...337..287E} found larger enhancements in the limb non-thermal broadening, with enhancement values up 8~\kms\ for lines formed close to the temperature of \ion{Si}{iv}. This result was interpreted as evidence of Alfv\'en waves passing through the TR. Most recently, \citet{Rao2022} studied the CLV of the non-thermal broadening of \ion{Si}{iv} using IRIS data, finding a limb non-thermal velocity of 20~\kms\ and a disk center velocity of 15~\kms, supporting the earlier \citet{1998A&A...337..287E} result. 

Section~\ref{sect:obs} presents the observational dataset used in the present work. Section~\ref{sect:results} describes the analysis method and gives the center-to-limb variation for the \ion{Si}{iv} emission line parameters. Finally, the results are summarized in Section~\ref{sect:summary}.

\section{Observation and Data Analysis}\label{sect:obs}

The Interface Region Imaging Spectrometer \citep[IRIS:][]{2014SoPh..289.2733D} provides high-resolution imaging and spectroscopic data at near ultraviolet (NUV) and far ultraviolet (FUV)  wavelengths. In the FUV spectrum are two strong lines at 1393.77~{\AA} and 1402.77~{\AA} that are due to \ion{Si}{iv}, an ion formed in the TR at a temperature around 80~kK. The 1402.77~{\AA} line is the subject of the present study.
%The \ion{Si}{iv} ion is formed in the TR at a temperature around 80~kK and gives two strong lines in the IRIS FUV spectrum at 1393.77~{\AA} and 1402.77~{\AA}, and the latter is the focus of the present study.
%An important pair of emission lines for IRIS are Si~{\sc iv} 1393.77~{\AA} and 1402.77~{\AA}, which are formed at 80~kK in the transition region and the latter line is the subject of the present study. 

%IRIS builds up spectroscopic images of the Sun by rastering a slit that has a spatial extent of 175\as. {\bf \color{red} The width of IRIS slit is 0.33\as. The coarse raster observations (i.e., 64 steps raster with step of 2\as) were being used to produce full-disk mosaics.} Data for the full solar disk requires observations at multiple consecutive spacecraft pointings and this is done regularly as part of the IRIS synoptic program \citep{Gunar2021}. 

IRIS has a slit of length 175\as and width 0.33\as, and spectroscopic images are built up through rastering. For the full-disk mosaic observations, 64-step rasters are used, and the step size is 2\as. Multiple, consecutive spacecraft pointings are required to observe the full solar disk. The mosaics are performed regularly as part of the IRIS synoptic program \citep{Gunar2021}. These data are ideal for studying CLV of emission line properties.

% The Interface Region Imaging Spectrometer \citep[IRIS:][]{2014SoPh..289.2733D} provides high-resolution imaging as well as spectroscopic observations. IRIS observes the Sun in the near ultraviolet (NUV) and far ultraviolet (FUV). This particular part of the solar atmosphere consists of various emission and absorption spectral lines which form in the different parts of the solar atmosphere. Si~{\sc iv} 1393.77~{\AA} and 1402.77~{\AA} are two important spectral lines that form in the solar transition region.\\
% IRIS captures FUV and NUV spectra from the small regions of the solar atmosphere. In addition to the small regions, the IRIS performs the campaign to capture the emission from full-disk Sun in various prominent spectral lines of FUV and NUV, namely, Si~{\sc iv} 1393.77~{\AA}, Si~{\sc iv} 1402.77~{\AA}, Mg~{\sc ii} k 2796.35~{\AA}, Mg~{\sc ii} h 2803.52~{\AA}, C~{\sc ii} 1334.53{\AA}, and C~{\sc ii} 1335.71{\AA}. For this particular project, we have used the IRIS full-disk mosaic in Si~{\sc iv} 1402.77~{\AA} that forms in the transition region.
In the present work we use the full-disk mosaic obtained between 13:09~UT on 24 September 2017 and 06:33~UT on 25 September. The mosaic comprised 185 coarse raster scans, each with an exposure time of 2~s. The IRIS team assembles the individual rasters into a single 3D data cube ($x$, $y$, $\lambda$) for each  emission line, and makes them available at the \href{https://iris.lmsal.com/mosaic.html}{IRIS Full Disk Mosaic website}.

In order to study the CLV within a coronal hole, it is necessary to clearly identify the coronal hole boundary within the IRIS full-disk mosaic image. 
For this purpose we used co-temporal 193~\AA\ images from the Atmospheric Imaging Assembly \citep[AIA:][]{Lemen2012} on the Solar Dynamics Observatory \citep[SDO:][]{2012SoPh..275....3P}. This channel is dominated by \ion{Fe}{xii} lines formed at around 1.5~MK that clearly reveal the coronal hole locations. By matching the observation times and spatial locations of IRIS, AIA 193~\AA\ images were assembled into a mosaic that shows the co-temporal coronal conditions for the IRIS mosaic. The left panel of Figure~\ref{fig:maps_qs} shows the  AIA 193~{\AA} mosaic, and the right panel shows the IRIS \ion{Si}{iv} mosaic. The coronal hole contours derived from the 193~\AA\ image are over-plotted on both full-disk images.
%Visually, we do see various regions (e.g., active regions, quiet Sun, and coronal holes) in this coronal image full-disk image of the Sun.\\

%IRIS has observed a total of 185 coarse raster scans from 24 September 2017 (t = 13:09:16~UT) to 25 September 2017 (t = 06:33:10~UT). All these coarse raster scans are 64-step raster observations. Finally, the full-disk mosaic spectroheliogram is produced using these 185 coarse raster observations. The exposure time of all these observations is 2 seconds.\\

% As we know that AIA/SDO observes the Sun in various wavebands (e.g., 4500~{\AA}, 1700~{\AA}, 1600~{\AA}, 304~{\AA}, 171~{\AA}, 131~{\AA}, 335~{\AA}, 193~{\AA}, and 94~{\AA}), and the various filters of AIA/SDO sample the emissions from the photosphere to the solar corona (see \citealt{Lemen2012}). Here, along with the IRIS mosaic, a pseudo-mosaics from AIA~193~{\AA} filter is also provided. Here, please note that AIA~193~{\AA} filter samples the emission from high-temperature plasma, i.e., coronal emission. The AIA 193~{\AA} pseudo-mosaics show the coronal emission at the time when IRIS was sampling the different portions of the solar disk\footnote{https://iris.lmsal.com/mosaic$\_$allin1.html}. The left panel of Figure~\ref{fig:maps_qs} shows the full-disk image of the Sun using AIA 193~{\AA} pseudo-mosaics. Visually, we do see various regions (e.g., active regions, quiet Sun, and coronal holes) in this coronal image full-disk image of the Sun.\\

The \ion{Si}{iv} 1402.77~{\AA} full-Sun mosaic  datacube was downloaded for the present study, and gives the spectral radiances for each spatial pixel of the solar disk. 
%The IRIS \ion{Si}{iv} mosaic datacube was extracted following the instructions at the \href{https://iris.lmsal.com/mosaic.html}{IRIS Full Disk Mosaic website}. 
Uncertainties on the spectral radiances were derived using the procedure contained in the SSW IDL routine \textsf{iris\_getwindata.pro}. \pry{The \ion{Si}{iv} 1402.77~{\AA} image shown in Figure~\ref{fig:maps_qs} was generated by first binning the mosaic by two and five pixels in the $x$ and $y$ directions, respectively. A single Gaussian function was then fit to the line at each spatial pixel using the IDL routine \textsf{gaussfit.pro}, and the image shows the radiances from the Gaussian fit functions.}

%To improve signal-to-noise, the data were binned by two and five pixels in the $x$ and $y$ directions, respectively. The \ion{Si}{iv} 1402.77~{\AA} line profile was fit with a single Gaussian using the IDL routine \textsf{gaussfit.pro}. The \ion{Si}{iv} image in Figure~\ref{fig:maps_qs} shows the radiance map obtained from fitting the \ion{Si}{iv} at each spatial pixel.

% We have used the IRIS mosaic to get the full-disk spectral intensity map from the Si~{\sc iv} 1402.77~{\AA} line. First of all, \textbf{the header and data array (a 3-D array: x, y, and $\lambda$) from IRIS mosaic spectroheliogram is extracted using readfits.pro, and this routine is available in the SSW. Further, to calculate the uncertainties (i.e., errors) in the intensities, we have followed the approach given in another routine of SSW ( iris$\_$getwindata.pro). Now, we have spectra as well as corresponding error arrays, and then}, we have binned the spectra to increase the signal-to-noise ratio. We have applied the binning of 2 and 5 pixels in the x and y directions, respectively. After this binning, we fitted the Si~{\sc iv} spectral line with a single Gaussian. \textbf{We have used gaussfit.pro routine for the Gaussian fit on the observed spectral profile, and this routine is available in the solar software (SSW).} Then we estimated the spectral intensity from each pixel of the full-disk mosaic. The full-disk spectral intensity image is shown in the right panel of Figure~\ref{fig:maps_qs}.

\section{Results}\label{sect:results}

In this section we obtain distributions of the intensity, Doppler velocity, and non-thermal line width of \ion{Si}{iv} 1402.77~\AA\ as a function of $\mu=\cos\,\theta$ for quiet Sun and coronal hole regions.  $\theta$ is the heliocentric angle ($\theta=0$ is disk center). 
\pry{The AIA and IRIS data used in this work provide spatial information in heliocentric coordinates. These were converted to $\mu$ values by using a solar radius of 957.3\as, obtained from the IDL routine \textsf{get\_rb0p} using the date 2017 September 24.}
The method for obtaining a reference wavelength for \ion{Si}{iv} 1402.77~\AA, which is needed to obtain Doppler velocities, is 
described in Section~\ref{section:rest_wave}. The selections of the QS and CH regions are discussed in Sections~\ref{section:qs_bound}, and the results are presented in Sections~\ref{section:clv_qs} and \ref{section:clv_chs}.

%This work is dedicated to understand the center-to-limb variations (CLV) in intensity, Doppler velocity, and line width deduced from quiet-Sun (QS) and coronal holes (CHs). \textbf{Firstly, in order to investigate the CLV in QS and CHs, the proper estimation of the rest wavelength of the Si~{\sc iv} line is required. Secondly, the proper selection of QS as well as CHs is necessary. Hence, let us start with the estimation of the rest wavelength of Si~{\sc iv} 1402.77~{\AA}. Then, we will describe the selection of the QS followed by the discussion on the CLV in intensity, Doppler velocity, and non-thermal velocity. We have followed the same approach for CHs, i.e., the selection of CHs, and then a discussion on the CLV in intensity, Doppler velocity, and non-thermal velocity.} 
\subsection{Estimation of Rest Wavelength} \label{section:rest_wave}

\pry{To derive the Doppler velocity CLV for \ion{Si}{iv} 1402.77~\AA\ it is necessary to have an absolute wavelength calibration reference. For the full-disk mosaic data used here, we assume that the 1402.77~\AA\ line has zero Doppler velocity at the limb ($\mu=0$). This is implemented by defining quiet Sun and coronal hole regions (Sects.~\ref{section:clv_qs} and \ref{section:clv_chs}) within the mosaics and plotting the centroids as a function of $\mu$. A straight line fit ($a+b\mu$) to the centroids is performed, and the parameter $a$ is set to the reference wavelength of the line (1402.770~\AA).{\color{black} Further details are given in Appendix~\ref{sect:append1} }. Therefore when the Doppler velocity is derived using the reference wavelength, the velocity at the limb is 0~\kms. }

\pry{The wavelength scale of IRIS is known to drift during an orbit (\href{https://www.lmsal.com/iris_science/doc?cmd=dcur&proj_num=IS0203&file_type=pdf}{IRIS Technical Note 20}), which potentially affects the mosaic Doppler velocity results. However, this effect is corrected in the IRIS level-2 data used here, hence the reference wavelength is assumed the same across the mosaic.}

\subsection{Selection of the QS and CHs} \label{section:qs_bound}
\subsubsection{Selection of the QS}

\begin{figure*}
    \includegraphics[trim = 3.5cm 1.0cm 3.8cm 1.0cm,scale=1.2]{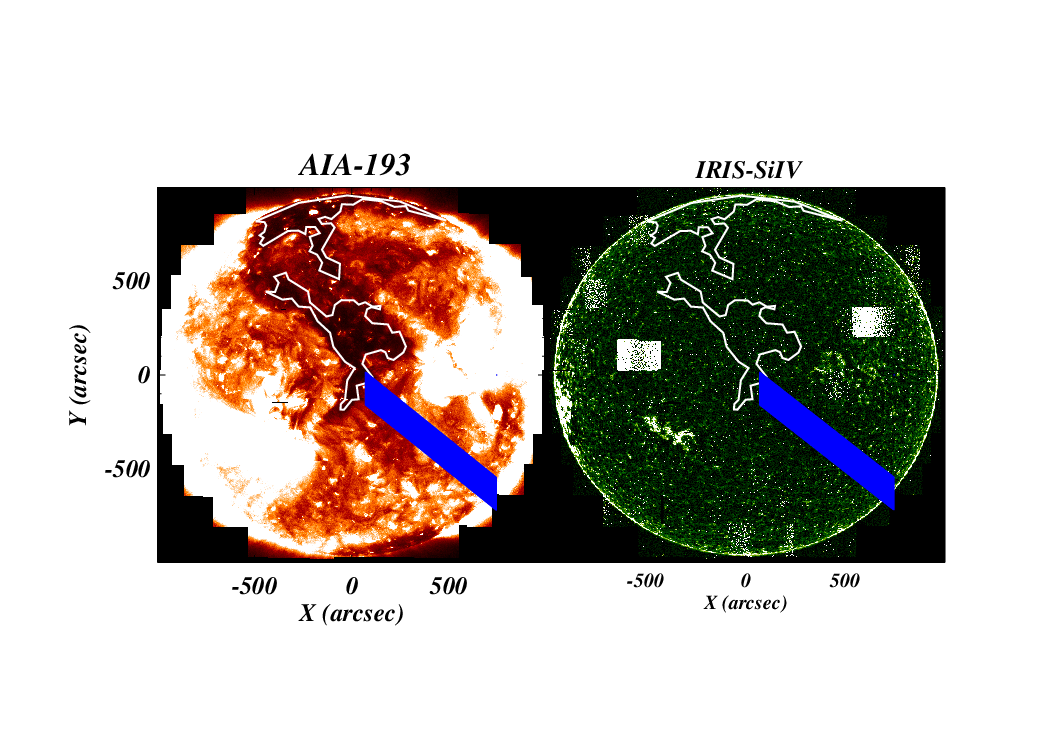}
    \caption{The right panel shows the IRIS full-disk mosaic image from the Si\,{\sc iv} 1402.77~{\AA} line. The left panel shows a mosaic image constructed from AIA 193~\AA\ images that were chosen to match the timing and pointing data from the IRIS mosaic. The QS region used for the CLV study is shown by a slanted blue patch that extends from the disk center towards the southwest limb of the Sun. White contours denote the two CHs that were used for the CLV study.}
    % The left panel shows an image of the Sun Left panel: It shows a full-disk pseudo image of the Sun. This pseudo AIA~193~{\AA} image is produced using several AIA~193~{\AA} observation files as per the area, date, and time of IRIS observations. The selected QS region is shown by a slanted blue patch that extends from the disk center towards the southwest limb of the Sun. In addition to QS region, we have also shown two CHs (i.e., white contours). Right-panel: IRIS Si~{\sc iv} 1402.77~{\AA} spectral intensity image with QS region (i.e., blue slanted patch) and two coronal holes (i.e., white contours).}
    \label{fig:maps_qs}
\end{figure*}

\pry{The south-west quadrant of the AIA 193~\AA\ image (Figure~\ref{fig:maps_qs}) is mostly free of CH and AR emission, and a diagonal band running from disk center to the limb was selected to derive the QS CLV. The IRIS \ion{Si}{iv} mosaic image was first binned by 5 pixels in the $y$-direction. For each of 339 $x$-pixels along the diagonal, 109 $y$-pixels centered on the diagonal were extracted resulting in an array of 339 x 109 spectra. This array is overplotted on Figure 2 in blue. The $\mu$ value was derived for each pixel within this array. The \ion{Si}{iv} line was fit to each spectrum, giving intensity, centroid, and line width data as a function of $\mu$.}
%%%%%%%%%%%%%%%%%%PK%%%%%%%%%%%%%%%%%%%%%%%%%%%%%%%%%%%%%%%%%%%%%%%%%%%%%%%
%The left panel of Figure~\ref{fig:maps_qs} shows the AIA-193~{\AA} image while the right panel shows IRIS Si~{\sc iv} spectral intensity full disk image of the Sun.  
%%%%%%%%%%%%%%%%%%%%%CLV in QS%%%%%%%%%%%%%%%%%%%%%%%%%%%%%%%%%%%%%%%%%%%%%%%%%%%%%%
%%%%%%%%%%%%%%%%%%%%%%%%%%%%%%%%%%%%%%%%%%%%%%%%%%%%%%%%%%%%%%
%The AIA~193~{\AA} intensity image (i.e., coronal intensity image) of the Sun shows the presence of various regions of the solar atmosphere, e.g., coronal holes (CHs), Quiet Sun (QS), and active-regions (ARs). 
%\textbf{Before selecting the QS region, to improve the signal-to-noise (SNR) ratio, we applied a binning of the 5 pixels (in the y-direction) on the full-disk mosaic. Then, we selected a QS path that extends from the disk center to the southwest limb of the Sun. Hence, this selected QS path covers the full $\mu$ range (i.e., from $\mu$ = 0.0 (disk center) to 1.0 (limb)). The selected QS path covers 70.0$"$ to 745.0$"$ in the x-direction and -70.0$"$ to -633.34$"$ in the y-direction. Here, please note that a total of 339 locations lie within the selected ranges of x- and y-directions. Further, we have taken 109 vertical locations at each position of the selected QS path. Hence the QS patch has a dimension of 339$\times$109 pixels. Finally,} the QS patch is displayed by a blue patch on the AIA~193~{\AA} and IRIS Si~{\sc iv} spectral intensity images (see; figure~\ref{fig:maps_qs}). Here, please note that the selected region is free from strong magnetic fields, and it has a relatively bright region. Hence, we have considered this region as the QS, and utilized it further to understand the CLV in QS.
\subsubsection{Selection of the CHs}
\pry{The white contours on Figure 2 show the locations of two coronal holes as obtained from the Heliophysics Event Knowledgebase \citep[HEK:][]{Hurlburt2012}. The HEK contains coronal positions generated from the Spatial Possibilistic Clustering Algorithm (SPoCA: \citealt{Verbeeck2014}) method. The SPoCA results within the HEK have a 6 hour cadence, and for each segment of the AIA and IRIS mosaic images, the nearest-in-time SPoCA contours were used. The IRIS mosaic image was binned by 5 pixels in the $y$-direction to improve signal-to-noise. Each spatial pixel of the IRIS mosaic that lies within the white contours of Figure 2 was extracted and a Gaussian fit performed to the \ion{Si}{iv} line. The $\mu$ value for each pixel was also computed, giving the intensity, centroid and line width as a function of $\mu$. Combined, the two coronal holes give almost complete coverage of $\mu$ from 0 to 1.}

\subsection{Gaussian fitting}

\pry{For each pixel of the QS and CH areas described in the previous section, a Gaussian function was fit to \ion{Si}{iv} 1402.77~\AA\ using the IDL \textsf{gaussfit} function. Following \citet{Rao2022} we have rejected fits with a reduced-$\chi^2$ value greater than five. Such fits can arise from data artifacts such as cosmic ray hits, and from non-Gaussian line profiles due to dynamic solar events, for example. The integrated line intensity is computed from the fit parameters, and the Doppler velocity is computed from the line centroid using the limb reference wavelength described in Section~\ref{section:rest_wave}. The Gaussian width parameter is converted to full-width at half-maximum (FWHM), and then input to the routine \textsf{eis\_width2velocity}, available in the \emph{SolarSoft} distribution. This routine subtracts the thermal and instrumental widths and outputs the non-thermal velocity \citep[see, e.g., Eq.~1 of][]{Rao2022}.
The instrumental width is 26~m\AA\ \citep{2014SoPh..289.2733D}, and a temperature of $\log\,T=4.87$ is used to compute the thermal width. The latter is the temperature of peak ionization of \ion{Si}{iv}. The width of the line's emissivity function is 0.19~dex, measured as the width at half-maximum of the function.
}

\subsection{Center-to-Limb Variations in QS} \label{section:clv_qs}

\begin{figure*}
	\includegraphics[trim = 6.5cm 1.0cm 6.8cm 0.0cm,scale=1.2]{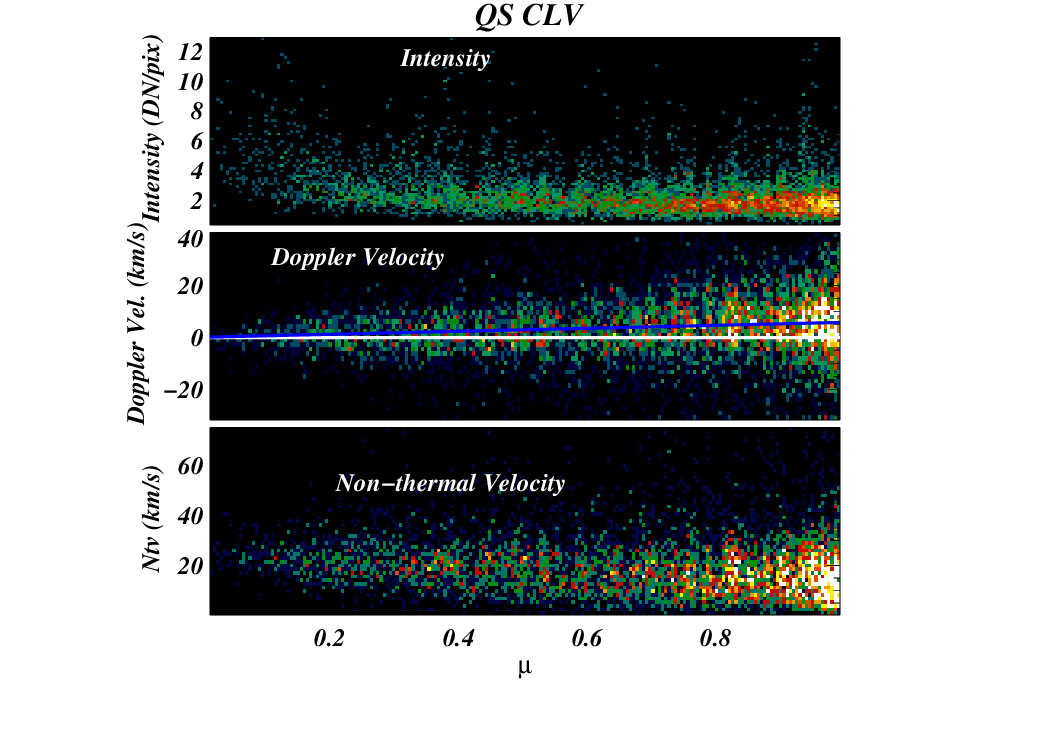}
    \caption{Two-dimensional histograms of spectral intensity (top panel), Doppler velocity (middle panel), and non-thermal velocity (bottom panel) plotted against $\mu$ for the QS region (the blue slanted patch in Figure~\ref{fig:maps_qs}). The blue line on the middle panel shows the fit to the Doppler velocity from Figure~\ref{fig:figure3}, and the white line denotes zero velocity.}
    \label{fig:clv_qs}
\end{figure*}

\begin{figure*}
	\includegraphics[trim = 6.5cm 1.5cm 6.8cm 0.0cm,scale=1.2]{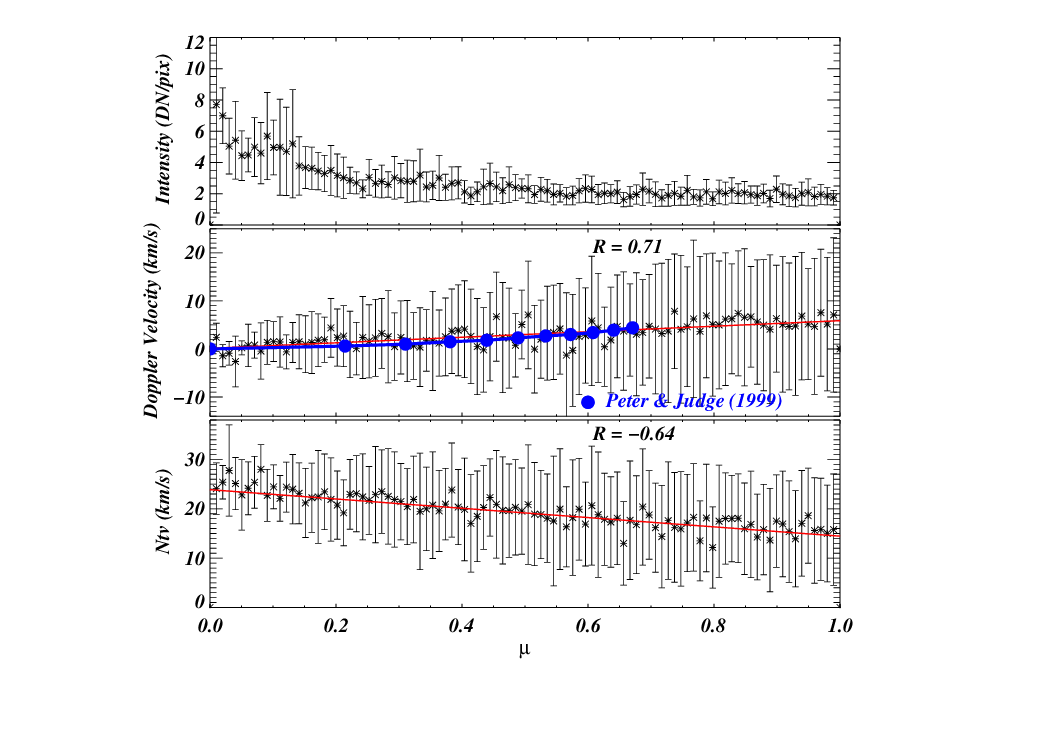}
    \caption{Plots of intensity (top), Doppler velocity (middle) and non-thermal velocity (bottom) averaged over 100 equally-spaced bins in $\mu$. Linear fits to the Doppler velocity and non-thermal velocity are over-plotted in red, and the fit parameters are given in Table~\ref{table:table1}. The Pearson correlation coefficients, $R$, are displayed on the middle and bottom panels.  The blue points on the middle panel are from \citet{PeterJudge1999}.}
    % We have shown $\mu$ vs. averaged intensity (top panel), Doppler velocity (middle panel), and line width (bottom panel) of the QS (see, blue slanted patch drawn on Figure~\ref{fig:maps_qs}). The red line is the line fit on each parameter with Pearson's coefficients of -0.6167 (intensity), 0.71644 (Doppler velocity), and -0.7265 (line width). All three parameters of the QS have CLV behavior.}
    % \caption{We have shown $\mu$ vs. averaged intensity (top panel), Doppler velocity (middle panel), and line width (bottom panel) of the QS (see, blue slanted patch drawn on Figure~\ref{fig:maps_qs}). The red line is the line fit on each parameter with Pearson's coefficients of -0.6167 (intensity), 0.71644 (Doppler velocity), and -0.7265 (line width). All three parameters of the QS have CLV behavior.}
    \label{fig:clv_qs_bin}
\end{figure*}

\pry{Figure~\ref{fig:clv_qs} shows 2D histograms of the intensity, Doppler velocity and non-thermal velocity as a function of $\mu$ as derived from the Gaussian fits to the QS pixels. In order to better display the center-to-limb trends, in Figure~\ref{fig:clv_qs_bin} we divide the $\mu$ range into 100 equally-spaced bins and plot the mean intensity, Doppler velocity and non-thermal velocity derived from the QS pixels within each bin. The error bars show the standard deviation of the values within that bin.}

\pry{The \ion{Si}{iv} intensity shows a small increase from disk center ($\mu=1$) to $\mu=0.2$, with a larger increase towards the limb. This is consistent with an increase in the line-of-sight column depth of the emitting plasma towards the limb. }

\pry{The Doppler velocity shows a clear trend of increasing redshift towards disk center (Figure 3) and a linear fit yields the parameters shown in Table~\ref{table:table1}. The Doppler shift at disk center is 5.2~\kms. \cite{PeterJudge1999} presented the CLV for  \ion{Si}{iv} 1402.77~\AA\ for a reduced range of $\mu$ values, and the fit to their data is overplotted in blue on the second panel of Figure~\ref{fig:clv_qs_bin}. The fit implies a redshift at disk center of $4.8\pm 1.1$~\kms, in excellent agreement with the value found here.}

\pry{The non-thermal velocity shows a clear decrease from the limb to disk center, and a linear fit yields the parameters shown in Table~\ref{table:table1}. The non-thermal velocity at the limb is 25~\kms\ and at disk center is 14~\kms. \citet{Rao2022} found values of around 20~\kms\ at the limb, and around 14{--}17~\kms\ near disk center using the \ion{Si}{iv} 1393.75~\AA\ line.}

%%%%%%%%%%%%%%%%%%%%%%%%%%%%PK%%%%%%%%%%%%%%%%%%%%%%%%%%%%%
%In this subsection, we will discuss the CLV in intensity, Doppler velocity, and line width parameters which are deduced from QS, i.e., from the blue slanted patch shown in figure~\ref{fig:maps_qs}.    
%%%%%%%%%%%%%%%%%%%%%%%%%%%%%%%%%%%%%%Figure: CLV for QS%%%%%%%%%%%%%%%%%%%%%%%%%
\begin{table}
\centering
\caption{Linear fit parameters ($a,b$) and Pearson correlation coefficients (PCC) for the Doppler velocity, $v_\mathrm{D}$, and non-thermal velocity, $\xi$, in the QS and CH regions. Please note that the error in each line fit parameter is the 1-$\sigma$ error.}
\begin{tabular}{ccccc}
\hline
Parameter & Region & $a$ & $b$ & PCC \\
\hline
$v_\mathrm{D}$ & QS & 0$\pm$0.33  & 5.74$\pm$0.57 & 0.71 \\
& CH & 0$\pm$0.32 & 4.84$\pm$0.54 & 0.68 \\
\noalign{\smallskip}
$\xi$  & QS & 23.8$\pm$0.65 & $-9.42$$\pm$1.12 & $-0.64$ \\
 & CH & 24.4$\pm$0.41 & $-14.8$$\pm$0.69 & $-0.91$ \\
\hline
\end{tabular}
\label{table:table1}
\end{table}

%%%%%%%%%%%%%%%%%%%%%%%%%%%%%%%%%%%%%%%%%%%%%%%%%%%%%%%%%%%%%%%%%%
 \subsection{Center-to-Limb Variations in CHs} \label{section:clv_chs}

\begin{figure*}
	\includegraphics[trim = 6.5cm 1.0cm 6.8cm 0.0cm,scale=1.2]{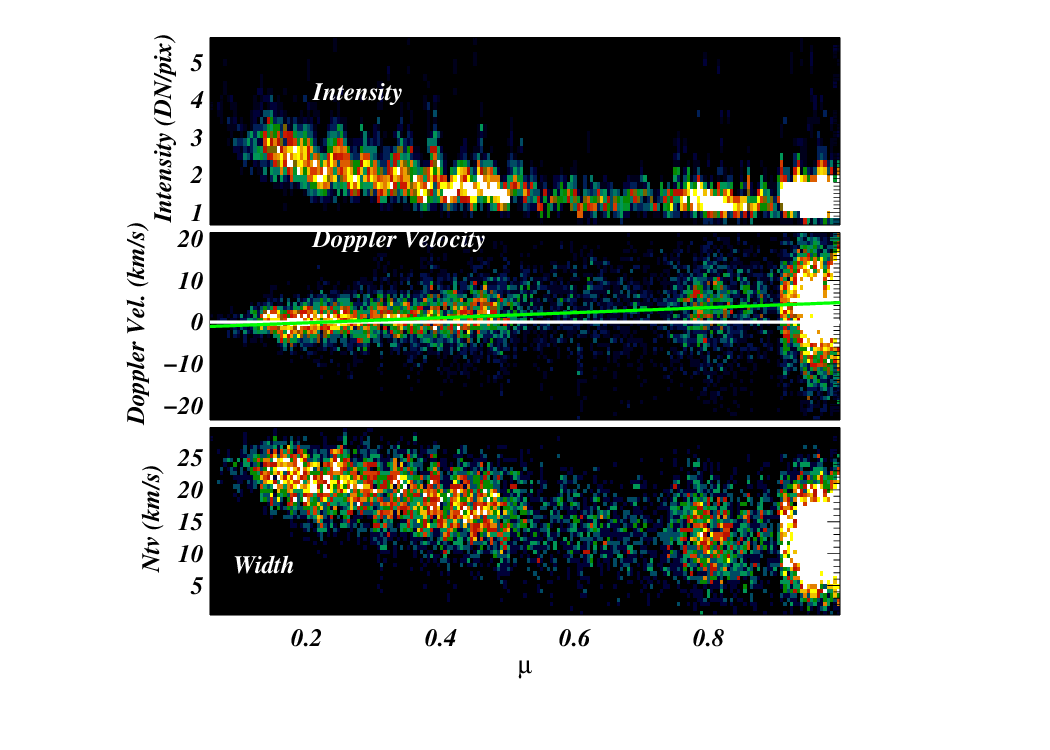}
    \caption{Two-dimensional histograms of spectral intensity (top panel), Doppler velocity (middle panel), and non-thermal velocity (bottom panel) plotted against $\mu$ for the CH region. The green line on the middle panel shows the fit to the Doppler velocity from Figure~\ref{fig:clv_qs_bin}, and the white line denotes zero velocity.}
%    \caption{We have deduced all the spectroscopic parameters (i.e., intensity, Doppler velocity, and line width) from both selected CHs, and then we also estimated $\mu$ values. We sorted intensity, Doppler velocity, and line width arrays as per $\mu$ value, and then we produced 2-D histogram for intensity (top panel), Doppler velocity (middle panel), and line width (bottom panel). It is clear that all three parameters show CLV. In the case of the Doppler velocity histogram, the white line is drawn on the zero velocity while green is the fitted trend on the Doppler velocity CLV.}
    \label{fig:figure3}
\end{figure*}

\begin{figure*}
	\includegraphics[trim = 6.5cm 1.5cm 6.8cm 0.0cm,scale=1.2]{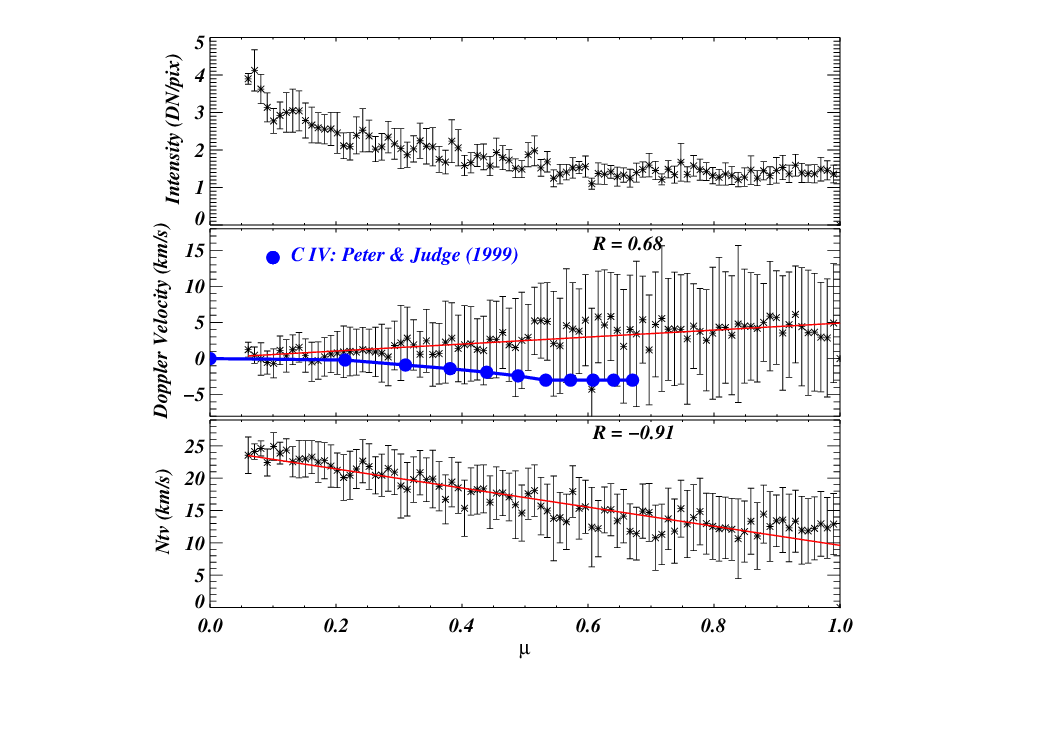}
    \caption{Same as Figure~\ref{fig:clv_qs_bin} but for the CH regions.}
    \label{fig:clv_ch_bin}
\end{figure*}

\pry{Figures~\ref{fig:figure3} and \ref{fig:clv_ch_bin} are directly analogous to Figures~\ref{fig:clv_qs} and \ref{fig:clv_qs_bin}, but for the CH data. Although there is almost complete coverage of the $\mu$ angles, there are relatively few pixels below $\mu=0.1$ and in the range $0.55\le \mu \le 0.75$. The behaviors of the intensity, Doppler velocity and nonthermal velocity are consistent with the QS dataset. The intensity shows a small increase to about $\mu=0.2$, and a larger increase towards the limb.
}

\pry{The Doppler velocity shows a clear linear relationship with $\mu$ (Figure~\ref{fig:clv_ch_bin}, middle panel), and the linear fit parameters are given in Table~\ref{table:table1}. The velocity at disk center is 4.8~\kms, very similar to the value found in the QS. \citet{PeterJudge1999} did not provide  Doppler velocity measurements in the coronal hole from \ion{Si}{iv}, but they did give values for \ion{C}{iv}, which is a slightly hotter ion ($\log\,T=5.05$). The fit from this work is over-plotted on the middle panel of Figure~\ref{fig:clv_ch_bin} as a blue line, and it can be seen that the line is actually blue-shifted. Our work contradicts this measurement.
}

\pry{The non-thermal velocity also shows a clear linear relationship with $\mu$ (Figure~\ref{fig:clv_ch_bin}, lower panel), and the linear fit parameters and Pearson correlation coefficient are given in Table~\ref{table:table1}. The velocity at the limb is very similar to that found for the QS, but the disk center value is lower, at 9.6~\kms. }
 
%%%Now, after locating the CHs (section~\ref{section:qs_bound}) and estimating the rest wavelength (section~\ref{section:rest_wave}), we extracted the Si~{\sc iv} spectra from all the locations lie in both CHs, i.e., all locations which lie within white contours as shown in the Figure~\ref{fig:maps_qs}. The radius of the Sun was estimated using get$\_$rb0p.pro for 2017 September 24 and using this radius and position coordinates, we have also estimated the $\mu$ value for each location that lies within both CHs, \textbf{i.e., $\mu$ value corresponding to each spectra of CHs. Further, the $\mu$ array is sorted in ascending order, and accordingly the spectra is also sorted. To improve the SNR ratio of the Si~{\sc iv} 1402.77~{\AA} spectral line, the sorted spectra array were binned by 5 locations.} The $\mu$ arrays is also binned by the same amount (i.e., 5 locations) to get the average $\mu$ value corresponding to the binned spectra. Finally, after this process, we have totals of 22645 spectra (i.e., 14161 from first CH and 21184 from second CH) along with corresponding $\mu$ values from both CHs.\\
%%%%%%%%%%%%%Figure: CLV Histogram for Coronal Hole CLV%%%%%

\section{Discussion and Conclusions}\label{sect:summary}
\pry{We have presented the first-ever measurement of the center-to-limb variation of Gaussian line-fit parameters  for a solar transition region line (\ion{Si}{iv} 1402.77~\AA) in a coronal hole. We have used data from an IRIS full-disk mosaic dataset that includes a coronal hole that extends to disk center. The line intensity increases towards the limb, consistent with an increasing plasma column depth due to line-of-sight effects. The Doppler velocity and non-thermal velocity show linear variations with $\mu$. The former is normalized to zero velocity at the limb, and displays an increasing redshift towards disk center. The disk center velocity is $4.8$~\kms. The non-thermal velocity  decreases from 24~\kms\ at the limb to 10~\kms\ at disk center.}

\pry{The same procedures were applied to a section of quiet Sun, and similar results were found. The redshift at disk center is 5.2~\kms, and the non-thermal velocity decreases from 25~\kms\ at the limb to 14~\kms\ at disk center.}

\pry{The quiet Sun Doppler velocity results are consistent with those from \citet{PeterJudge1999}, but the coronal hole results are significantly different. \citet{PeterJudge1999} found blue-shifted emission in a coronal hole from \ion{C}{iv}, which is a slightly hotter ion than \ion{Si}{iv}. \citet{Rao2022} recently measured non-thermal velocities in the quiet Sun using the IRIS \ion{Si}{iv} 1393.75~\AA\ line, and they found similar values at disk center but slightly lower velocities of 20~\kms\ at the limb.}

\section*{Acknowledgements}
PRY acknowledges support from the GSFC Internal Scientist Funding Model competitive work package program. IRIS is a NASA Small Explorer mission developed and operated by LMSAL with mission operations executed at NASA Ames Research Center and major contributions to downlink communications funded by ESA and the Norwegian Space Centre. We also acknowledge the data from SDO/AIA. 
%%%%%%%%%%%%%%%%%%%%%%%%%%%%%%%%%%%%%%%%%%%%%%%%%%
\section*{Data Availability}
We used spectroscopic data from IRIS and imaging observations (i.e., AIA~193~{\AA} mosaic) from AIA/SDO instruments, and the complete set of data is available 
at \url{https://www.lmsal.com/solarsoft/irisa/data/level2_compressed/2017/09/24/20170924Mosaic/}. 


\bibliographystyle{mnras}
\bibliography{qs_ch_clv_r1}

%\begin{thebibliography}{99}
%\end{thebibliography}

%%%%%%%%%%%%%%%%%%%%%%%%%%%%%%%%%%%%%%%%%%%%%%%%%%

%%%%%%%%%%%%%%%%% APPENDICES %%%%%%%%%%%%%%%%%%%%%

\appendix
\section{Rest Wavelength Estimation: Centriod vs $\mu$} 
\label{sect:append1}
 We have estimated the mean and standard deviation of the centroid at 100 $\mu$ values between 0.0 to 1.0 (disk center to limb). Please see section~\ref{section:rest_wave} for more details.
\setcounter{figure}{0}
\renewcommand\thefigure{\thesection.\arabic{figure}}
%%%%%%%%%%%%%Centriod VS CLV %%%%%%%%%%%%%%%%%%%%%%%%%%%
\begin{figure*}
	\includegraphics[trim = 2.0cm 0.0cm 2.0cm 0.0cm,scale=1.0]{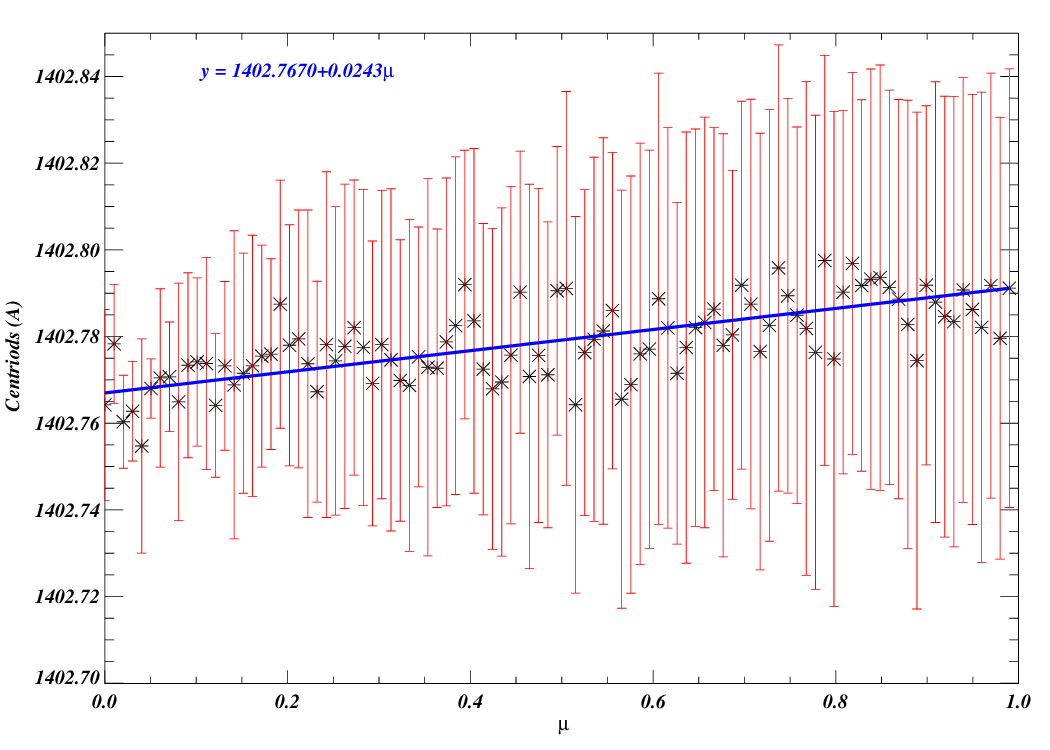}
    \caption{The figure shows variations of centroid with $\mu$ values. The errors corresponding to each centroid are shown in red color. The blue line is the linear fit on the centroid, and the fitted line y = 1042.7670+0.0243$\mu$. Hence, the centroid at the limb is 1402.7670~\AA(or ~1402.77~{\AA}) which is considered as the rest wavelength.}
    \label{fig:append1}
\end{figure*}
%%%%%%%%%%%%%%%%%%%%%%%%%%%%%%%%%%%%%%%%%%%%%%%%%%%%%%%%%%%%
The mean centroid versus $\mu$ is fitted with a straight line (see blue line in figure~\ref{fig:append1}). The fitted line is as follows: $y$ = 1402.7670+0.0243$\mu$. Hence, the centroid at first $\mu$ value (i.e., at solar limb) is 1402.7670~{\AA}, and it is considered as the rest wavelength of Si~{\sc iv} spectral line to calculate the Doppler velocity.  

% Don't change these lines
\bsp	% typesetting comment
\label{lastpage}
\end{document}